\documentstyle[aps]{revtex}
\newcommand{\bd}{\begin{displaymath}}
\newcommand{\ed}{\end{displaymath}}
\newcommand{\be}{\begin{equation}}
\newcommand{\ee}{\end{equation}}
\newcommand{\barr}{\begin{eqnarray}}
\newcommand{\earr}{\end{eqnarray}}
\newcommand{\barrr}{\begin{eqnarray*}}
\newcommand{\earrr}{\end{eqnarray*}}

\title{Is the purely biquadratic spin 1 chain\\
always massive?}
\author{\em Giuseppe Albertini\thanks{E-mail 
albert@osfmi.mi.infn.it}\\
\em Istituto di Fisica, Universita' Statale\\
\em via Celoria 16, 20133 Milano (Italy)} 
\date{December 22, 2000}

\begin{document}

\maketitle

\begin{abstract}
It is shown that the $sl(2)_q$-invariant open antiferromagnetic $XXZ$ 
spin chain with a boundary field has a gapless sector in the thermodynamic 
limit when its length is odd. Owing to a Temperley-Lieb equivalence of the
spectra, the same conclusion is drawn for the purely biquadratic spin 1 
chain with open boundaries and odd length. 
\end{abstract}



IFUM-674-FT 

\newpage

\noindent
The bilinear-biquadratic spin 1 quantum chain
\be\label{e1}
H = \sum_{n=1}^{L} \cos \theta {\bf S}_n \cdot {\bf S}_{n+1} + \sin \theta 
({\bf S}_n \cdot {\bf S}_{n+1})^2
\ee
has been the object of very intense investigation in the last twenty years.
One of the goals has been to determine the properties of the ground state
and the nature of the low-lying excitations in the thermodynamic limit. 
In particular, whether they form a continuum with the ground state 
(critical chain) or whether they are separated by a finite gap (massive chain).
Haldane's prediction \cite{Ha} that the Heisenberg antiferromagnet 
($\theta = 0$)
should be massive for integer spin has been thouroughly verified by a 
variety of numerical methods \cite{WH}. The AKLT chain 
$(\tan \theta =\frac{1}{3})$ has been rigorously proven to have a valence bond
ground state and a nonzero gap \cite{AKLT}.
Other cases of (\ref{e1}) have been
analyzed in depth due to their integrability. They are the critical $SU(3)$-
invariant Sutherland spin chain ($ \theta = \pi /4$), solved by nested
Bethe-ansatz \cite{Su}; the Babujian-Takhtajian chain ($\theta = -\pi/4$), 
also solvable by Bethe-ansatz and also gapless \cite{Ta,Bb}. The third case is
the purely biquadratic spin chain ($ \theta = -\pi/2 $), which, after some
controversy, has been found to have a gap $\Delta \simeq 0.1731788$ \cite{Kl}; 
an identical gap has been revealed in the same chain, but this time with 
free boundary conditions
\be\label{e2}
H_{{\rm bQ}} = - \sum_{n=1}^{L-1} ({\bf S}_n \cdot {\bf S}_{n+1})^2
\ee
and $L$ {\em even} \cite{BB}.

Now, at least for free boundary conditions, it can be shown that (\ref{e2}) 
has a band of gapless excitations when $L$ is {\em odd} and,
of course, $L \rightarrow \infty$. This points to the fact that 
antiferromagnetic quantum spin chains can in principle have very different 
properties according to the parity of their length.

To prove this statement, one observes that the spectrum of (\ref{e2}) can be 
mapped \cite{BB}
into the spectrum of the $sl(2)_q$-invariant open antiferromagnetic 
$XXZ$ chain with a boundary field
\be\label{e3}
H_{XXZ} = -\frac{1}{2} \sum_{n=1}^{L-1}(\sigma_{n}^x \sigma_{n+1}^x +
\sigma_{n}^y \sigma_{n+1}^y - \cosh \gamma \sigma_{n}^z \sigma_{n+1}^z )
+ \frac{\sinh \gamma}{2} (\sigma_1^z - \sigma_L^z)
\ee
The equivalence holds when $\cosh \gamma =\frac{3}{2}$, $\sinh \gamma =
\frac{\sqrt{5}}{2}$. At this point both (\ref{e2}) and (\ref{e3}) can be 
written as sums over the generators $\{e_n\}_{n=1}^{L-1}$ of the 
Temperley-Lieb algebra \cite{BB,TL}
\barr
&&e_n^2 = 3e_n \hspace{1.5cm} e_n e_{n \pm 1} e_n = e_n \hspace{1.5cm} 
[e_n,e_{n'}]=0  \ \ \ |n-n'|\geq 2 \label{e4} \\
&&H_{bQ} = H_{XXZ}- \frac{7}{4}(L-1)  \label{e5}
\earr
Eq. (\ref{e5}) must be understood as a statement about the spectra. The two
representations of (\ref{e4}), appearing in (\ref{e2}) and (\ref{e3}), are of 
course different since
the two Hamiltonians live in Hilbert spaces of different dimensions, but their
spectra should differ only in their multiplicities \cite{BB,Ba}
(further comments on this point will be given at the end). Similar mappings, 
via Temperley-Lieb algebra, have shown to be very fruitful in studying 
statistical
mechanics models and related quantum spin chains, one notable example being
the Potts model \cite{Ba,HK}.

The spin chain (\ref{e3}) has been solved by means of the coordinate 
Bethe-ansatz (BA) \cite {ABBBQ}. In fact, it was relying on the Temperley-Lieb 
mapping and the numerical as well as analytical solution of the relevant BA 
equations that the ground
state energy and the gap of (\ref{e2}) were computed in \cite{BB}. As in all 
BA solvable systems, the spectrum of (\ref{e3}) is expressed in terms of 
rapidities $\{\alpha_1, \ldots \alpha_n \}$ that solve a set of coupled 
equations. In the case at hand, they are
\barr
&&\frac{1}{\pi}\Theta(\alpha_j;\frac{\gamma}{2})- \frac{1}{2\pi L}
\sum_{k=1, k \neq j}^{n} \Big( \Theta(\alpha_j- \alpha_k ; \gamma) + \Theta
(\alpha_j+ \alpha_k ; \gamma) \Big) = \frac{I_j}{L} \hspace{1cm} 
j=1 \ldots n \hspace{1cm} \label{e6} \\
&&\Theta(\alpha ; x) \stackrel{def}{=} -i\ln \left[ \frac{\sinh(x+
\frac{i\alpha}{2})}{\sinh(x-\frac{i\alpha}{2})} \right] 
= 2\arctan(\tan \frac{\alpha}{2} \coth x) \label{e7}
\earr
Solutions are labelled by the set of {\em positive integers} 
$\{I_j \}_{j=1}^{n}$.
The branch cut in (\ref{e7}) is chosen to make $\Theta(\alpha;x)$ a 
differentiable, increasing function for $\alpha$ real, with $\Theta(-\alpha;x) 
= -\Theta(\alpha,x)$
and $\Theta(\alpha +2\pi ;x) = \Theta(\alpha ;x) + 2\pi$. Only positive
real rapidities in the $(0, \pi)$ interval need to be considered here. 
Given a solution of (\ref{e6}), the corresponding energy and spin $S_{tot}^z =
\frac{1}{2} \sum_{j=1}^L \sigma_j^z$ are \cite{ABBBQ}
\bd
E=\frac{1}{2}(L-1)\cosh \gamma -2 \sinh \gamma \sum_{j=1}^{n} \Theta' 
(\alpha_j ;\frac{\gamma}{2}) \hspace{1.5cm} S_{tot}^z = \frac{L}{2}-n
\ed
It's now crucial to determine the range of allowed vacancies for the set 
$\{I_j\}_{j=1}^{n}$. This is done by rewriting (\ref{e6}) in terms of the 
counting function \cite{FT}, defined as
\barr
&&Z_L(\alpha) \stackrel{def}{=} \frac{1}{\pi} \Theta (\alpha ; 
\frac{\gamma}{2}) +\frac{1}{2\pi L} \Big[ \Theta (\alpha ;\gamma) 
+ \Theta (2\alpha;\gamma) \Big] -\frac{1}{2\pi L} \sum_{k=-n}^{n} 
\Theta(\alpha -\alpha_k; \gamma) \hspace{1cm} \nonumber \\
&&Z_L(\alpha_j)= \frac{I_j}{L} \hspace{2cm} j=-n,-n+1, \ldots ,
n-1,n \label{e9}
\earr
Rapidities have been doubled by reflection through $\alpha =0$ \cite{ABBBQ}:
$\{\alpha_1, \ldots, \alpha_n \ \ | \ \ \alpha_j >0 \} 
\rightarrow \{\alpha_{-n}, \ldots,
\alpha_{-1}, 0, \alpha_1, \ldots, \alpha_n \ \ | \ \ \alpha_{-j} 
=- \alpha_{j}\}$,
so that (\ref{e9}) is completely equivalent to (\ref{e6}). 
Since the rapidity range is $(0, \pi)$, or $(-\pi, \pi)$ after reflection, 
and the counting function is monotonically increasing, the largest allowed 
$I_j$ is read from $L Z_{L}(\pi)$, or, conceptually more correctly but 
with equal numerical result
\be\label{e10}
\lim_{\alpha_j \rightarrow \pi} L Z_{L}(\alpha_j) = L-n+1
\ee
The limit does not depend on the numerical value of the remaining 
$\alpha_k$, $k \neq j$.
Eq. (\ref{e10}) would suggest $I_{max} = L-n+1$. Actually this value is 
forbidden because the wave-function vanishes when one rapidity is 
$\pi$ ($\alpha$ is related to the variable $k$ of \cite{ABBBQ} by $e^{ik} = 
(e^{i\alpha}-e^{-\gamma})/(1-e^{i\alpha -\gamma})$). Hence the largest 
possible integer is
\be\label{e10a}
I_{max} = L-n
\ee
If $L$ is even, $n=\frac{L}{2}$ for the ground state (sector $S_{tot}^z =0$)
and there's exactly 
$\frac{L}{2}$ vacancies for $\frac{L}{2}$ positive integers. But if $L$ is
odd, the lowest energy states have $n=\frac{L-1}{2}$ (sector $S_{tot}^{z} = 
1/2$) and there are $\frac{L+1}{2}$ vacancies for $\frac{L-1}{2}$ integers, 
hence one hole. Altogether a band of $\frac{L+1}{2}$ configurations, specified
by a sequence of closely packed integers with one hole $I^{(h)}$, $1 \leq 
I^{(h)} \leq \frac{L+1}{2}$. The ground state for finite $L$ has $I^{(h)}
= \frac{L+1}{2}$, that is at the edge of the band,
as it will be proven shortly.

Denote by $\{\alpha_j\}_{j=1}^n$ the rapidities of the state with $I^{(h)} =
\frac{L+1}{2}$, by $\{\alpha'_j\}_{j=1}^{n}$ the rapidities of any other 
state in the band ($1 \leq I^{(h)} \leq \frac{L-1}{2}$) and by $Z_L^{(0)} 
(\alpha)$ and $Z_{L}^{(1)} (\alpha)$ the 
relevant counting functions. One defines a hole rapidity $\alpha^{(h)}$ by
\bd
Z_{L}^{(1)}(\alpha^{(h)})=\frac{I^{(h)}}{L}
\ed
This number does not belong to the set $\{\alpha'_j\}_{j=1}^n$, but it is
convenient to include it, adding and subtracting its contribution to the
counting function, which reads after reflection
\barrr
&&Z_L^{(1)}(\alpha) = \frac{1}{\pi} \Theta (\alpha ; 
\frac{\gamma}{2}) +\frac{1}{2\pi L} \Big[ \Theta (\alpha ;\gamma) 
+ \Theta (2\alpha;\gamma) \Big] - \frac{1}{2\pi L} \sum_{k=-n-1}^{n+1} 
\Theta(\alpha -\alpha_{k}'; \gamma) + \\
&&\frac{1}{2\pi L} \Big( \Theta(\alpha - \alpha^{(h)}; \gamma) +
\Theta (\alpha + \alpha ^{(h)}; \gamma) \Big)
\earrr
so that
\be\label{e11}
Z_{L}^{(0)}(\alpha_k) - Z_{L}^{(1)}(\alpha'_k) = 0 \hspace{2cm} -\frac{L-1}{2}
\leq k \leq \frac{L-1}{2}
\ee
For these values of $k$, define $\delta \alpha_k = \alpha'_k - \alpha_k =
O(1/L)$. The two (after reflection) unpaired edge
rapidities are handled separately. Since $\{\alpha_{j}\}_{j=-n}^n$ fill the
interval $(-\pi, \pi)$ in the thermodynamic limit \cite{YY}, one can set 
$\alpha'_{\frac{L+1}{2}} = \pi + o(1)$
and $\alpha'_{- \frac{L+1}{2}} = - \pi +o(1)$ as $L \rightarrow \infty$.
The next steps follow a well-established method for handling the BA 
equations \cite{Ga,Li}. Defining
$\Delta \alpha_k = \alpha_{k+1} - \alpha_{k} = O(1/L)$ and the backflow
$\delta(\alpha) = \lim_{L \rightarrow \infty} \frac{\delta \alpha_k}{\Delta
\alpha_k}=O(1)$,
it is found that the terms $O(1)$ in (\ref{e11}) cancel and the 
terms $O(1/L)$ yield
\barr
&&\delta(\alpha) + \frac{1}{2 \pi} \int_{- \pi}^{\pi} d\beta \Theta'(\alpha -
\beta; \gamma) \delta(\beta) = \frac{1}{2\pi} \Big[ \Theta(\alpha - \pi; 
\gamma)+ \Theta(\alpha + \pi; \gamma) \nonumber \\
&&- \Theta( \alpha - \alpha^{(h)}; \gamma) - \Theta(\alpha + \alpha^{(h)}; 
\gamma) \Big] \label{e12} \\
&&\Delta E(\alpha^{(h)}) = \lim_{L \rightarrow \infty} \Big( E^{(1)} 
(\alpha^{(h)})-E^{(0)} \Big) = - \sinh \gamma \int_{- \pi}^{\pi} d\alpha 
\Theta''(\alpha; \frac{\gamma}{2}) \delta(\alpha) \nonumber \\
&&+ 2 \sinh \gamma \Big( \Theta' (\alpha^{(h)}; \frac{\gamma}{2}) - 
\Theta'(\pi; \frac{\gamma}{2}) \Big) \label{e13}
\earr
It is straightforward to solve (\ref{e12}) by Fourier transform. It is 
actually better to reduce it to an equation for $\delta'(\alpha)$ by 
differentiating and then
using $\delta(\pi) = \delta (- \pi)$ which follows from (\ref{e12}) itself. 
$\delta' (\alpha)$ is substituted after (\ref{e13}) has been integrated by 
parts. The result is
\barr
&&\Delta E(\alpha^{(h)}) = \epsilon(\alpha^{(h)}) - \epsilon(\pi) 
\hspace{2cm} 0 < \alpha^{(h)} < \pi \label{eadd} \\
&& \epsilon(\alpha) = 2 \sinh \gamma \sum_{n=- \infty}^{+ \infty} 
\frac{e^{-in\alpha}}{2 \cosh \gamma n} = 2 \sinh \gamma \,
\frac{K(k)}{\pi}{\rm dn}(\frac{K(k)\alpha}{\pi};k)  \nonumber \\
&&\frac{K'(k)}{K(k)} = \frac{\gamma}{\pi} \nonumber
\earr
where $K(K')$ is the real (imaginary) quarter period and $k$ the modulus
of the elliptic dn function \cite{Er}. This shows that $\Delta E
(\alpha^{(h)}) > 0$ and it goes to 0 when $\alpha^{(h)} \rightarrow \pi$, as
claimed. Moreover, the whole band found
for $L$ odd in $S^z_{tot} = 1/2$ is degenerate with a twin band in the
sector $S^z_{tot} = -1/2$. In fact, despite the boundary field, spectra in
the sectors $S^z_{tot} =S$ and $S^z_{tot} = -S$ coincide because of the
symmetry implemented by the unitary operator $I \prod_{n=1}^{L}\sigma_{n}^x$, 
where $I{\bf \sigma }_nI = {\bf \sigma }_{L+1-n}$. 

Excitations of the $XXZ$ chain are sometimes called spinons. When $L$ is even
they appear in even number \cite{FT,BVV}, hence they are assigned a spin
$S^z=1/2$. Eq.(\ref{eadd}) shows that, when $L$ is odd, the ground state 
sector must
contain one spinon which can be in a whole band of dynamical states with
varying energy. Such states cannot be labelled by a linear momentum 
because (\ref{e3}) is not traslationally invariant.

The case $L$ even is known. In the sector $S^z_{tot}=1$ there are 
$n=\frac{L}{2}-1$ rapidities; this, from (\ref{e10a}), implies 
$\frac{L}{2}+1$ vacancies for $\frac{L}{2}-1$ integers, hence two holes 
leading to
\bd
\Delta E(\alpha^{(h,1)}, \alpha^{(h,2)}) = \epsilon (\alpha^{(h,1)}) +
\epsilon(\alpha^{(h,2)}) \hspace{2cm} 0 < \alpha^{(h,1)} < \alpha^{(h,2)} < \pi
\ed
This coincides with the well know result for the periodic $XXZ$ chain, the 
only difference being that now hole rapidities lie in the restricted range
$(0, \pi)$.
At $\cosh \gamma = \frac{3}{2}$ the gap is of course $2 \epsilon(\pi) 
\simeq 0.1731788$ \cite{BB}. 

Other quantities have a different limit for $L$ odd or even, one example being
the surface energy $e^{(s)}$, defined by
\bd
E^{(0)}(L) = e_0L+e^{(s)}+o(1) \hspace{2cm} L \rightarrow \infty
\ed
For $L$ even it has been calculated in \cite{BH} through an analysis of finite
size corrections. Since the $XXZ$ periodic chain has the same $e_0 =
\frac{\cosh \gamma}{2} - \sinh \gamma (1+ 4 \sum_{n=1}^{+ \infty} \frac{1}
{1+e^{2n\gamma}})$ \cite{YY} but no surface energy, 
$e^{(s)}$ can also be found from
\bd
2E^{(0)}(L)-E^{(0,p)}(2L)= 2e^{(s)} +o(1) \hspace{2cm} L \rightarrow \infty
\ed
where the label ``p'' stands for periodic boundary conditions. 
This is an order
1 calculation which bypasses the intricacies of finite size calculations.
For the periodic chain of length $2L$, the ground state has $n=L$ \cite{YY,Ga}
\barrr
&&Z^{(0,p)}_{2L}(\alpha) = \frac{1}{\pi}\Theta(\alpha;\frac{\gamma}{2}) -
\frac{1}{2 \pi L}\sum_{k=1}^{L}\Theta(\alpha -\alpha_k;\gamma) \\
&&Z^{(0,p)}_{2L}(\alpha_j)=\frac{I_j}{L} \\
&&E^{(0,p)}(2L) = L\cosh \gamma -2 \sinh \gamma \sum_{k=1}^L \Theta'(\alpha_k;
\frac{\gamma}{2})
\earrr
Consider, when $L$ is odd, the lowest-lying state in the previously discussed
band of the open spin chain. Its $(L-1)/2$ rapidities, which become exactly
$L$ after reflection, will now be denoted 
$\{\alpha_j'\}_{j=-(L-1)/2}^{(L-1)/2}$, whereas
$\{\alpha_j\}_{j=-(L-1)/2}^{(L-1)/2}$ will be the $L$ 
symmetrically distributed ground state rapidities of the periodic chain 
of length $2L$. Hence
\bd
Z^{(0)}_{L}(\alpha_j') = Z^{(0,p)}_{2L}(\alpha_j) =\frac{j}{L}
\hspace{2cm} -\frac{L-1}{2} \leq j \leq \frac{L-1}{2}
\ed
The relevant equations in the thermodynamic limit are now
\barrr
&& \delta(\alpha) + \frac{1}{2\pi}\int_{-\pi}^{\pi}d\beta \Theta'(\alpha -
\beta;\gamma) \delta(\beta) = - \frac{1}{2\pi} \Big[ \Theta(\alpha; \gamma) +
\Theta (2\alpha ; \gamma) \Big] \\
&& 2e^{(s)} = - \cosh \gamma +2 \sinh \gamma\, \Theta' (0; \frac{\gamma}{2})
-2 \sinh \gamma \int_{-\pi}^{\pi} d\alpha \Theta''(\alpha; \frac{\gamma}{2})
\delta(\alpha)
\earrr
for the backflow $\delta(\alpha) = \lim_{L \rightarrow \infty} \frac{\delta
\alpha_{j}}{\Delta \alpha_j}$ (as before $\delta \alpha_j =\alpha'_j - 
\alpha_j$,
$\Delta \alpha_j =\alpha_{j+1}-\alpha_{j}$). The upshot is
\bd
e^{(s,odd)}=- \frac{\cosh \gamma}{2} +\sinh \gamma \left( 1+2\sum_{k=1}
^{+\infty}\frac{2-e^{-2k\gamma}}{\cosh 2k\gamma} - \sum_{k=0}^{+\infty}
\frac{2}{\cosh(2k+1)\gamma} \right)
\ed
For $L$ even, the calculation is slightly more involved, because now the
numbers $I_j$ are half-odd for the periodic chain of length $2L$, yet it
can be carried out along the same lines, leading to
\bd
e^{(s,even)} = -\frac{\cosh \gamma}{2} +2\sinh \gamma \, \sum_{k=1}
^{+\infty} \frac{1-e^{-2k \gamma}}{\cosh 2k \gamma}
\ed
A similar method was used in \cite{KS} for general boundary fields but,
again, for $L$ even only.
When $\cosh \gamma =3/2$, the biquadratic chain surface energy is,
from (\ref{e5}),\, $e^{(s)}_{{\rm bQ}} = e^{(s)}+7/4$, or
\bd
e^{(s,odd)}_{{\rm bQ}} \simeq 1.7415986 \hspace{2cm}
e^{(s,even)}_{{\rm bQ}} \simeq 1.6550092
\ed
The second value coincides with that in \cite{BH}.

I have checked numerically, for small chains of odd length, that:
1. eq. (\ref{e6}) admit a solution for any value of $I^{(h)}$ within the 
allowed range; 2. the corresponding energies show up in the spectrum of
(\ref{e3}), as found by direct diagonalization, and $I^{(h)}=\frac{L+1}{2}$ 
yields the ground state energy; 3. the same eigenvalues show up in the 
spectrum of (\ref{e2}), also obtained by direct diagonalization. 
The last check was carried out in \cite{BB} too (perhaps only for $L$ even).

Actually, the eigenvalue problem of (\ref{e2}) has a direct BA solution 
that does not rely on the mapping (\ref{e4},\ref{e5}) \cite{KL}, but 
the relevant BA equations
do not seem to have been studied in detail. This would provide a further check.
Finally, it should be pointed out that, very recently, peculiarities of
$L$ odd have been noticed also for the periodic $XXZ$ chain in the critical
regime \cite{St}.

\end{document}